\documentclass[a4paper,preprint,aps,floatfix,showpacs]{revtex4}

\usepackage{dcolumn}
\usepackage{graphicx}
\usepackage{amsfonts}
\usepackage{amsmath,bm,amssymb}
\usepackage{comment}
\usepackage[usenames]{color}

\begin{document}

\title{Engineering and probing topological properties of Dirac semimetal films by asymmetric charge transfer}
\author{John W. Villanova, Edwin Barnes, and Kyungwha Park}
\email{kyungwha@vt.edu}
\affiliation{Department of Physics, Virginia Tech, Blacksburg, Virginia 24061, USA}

\date{\today}


\begin{abstract}

Dirac semimetals (DSMs) have topologically robust three-dimensional Dirac (doubled Weyl) nodes with Fermi-arc states. In heterostructures involving DSMs, charge transfer occurs at the interfaces, which can be used to probe and control their bulk and surface topological properties
through surface-bulk connectivity. Here we demonstrate that despite a band gap in DSM films, asymmetric charge transfer at the surface enables one
to accurately identify locations of the Dirac-node projections from gapless band crossings and to examine and engineer properties of the topological Fermi-arc surface states connecting the projections, by simulating adatom-adsorbed DSM films using a first-principles method with an effective model. The positions of the Dirac-node projections are insensitive to charge transfer amount or slab thickness except for extremely thin films. By varying
the amount of charge transfer, unique spin textures near the projections and a separation between the Fermi-arc states change, which can be observed
by gating without adatoms.


\end{abstract}


\maketitle



Dirac and Weyl semimetals (DSMs and WSMs) are topological metals where bulk conduction and valence bands touch at an even number of
momentum ${\mathbf k}$ points with Dirac dispersion at or near the Fermi level ($E_F$)
\cite{WAN11,YOUN12,WANG12,LIU14,WENG15,XU15_Dirac,XU15_Weyl,LIU14_CdAs,WANG13_CdAs}. The band touching points are referred to as
Weyl nodes and they are topologically protected. Depending on the positions of the Weyl nodes in the bulk, on particular surfaces the
projections of the Weyl nodes of opposite chirality are connected by Fermi-arc states which can give rise to exotic transport
phenomena. Weyl nodes cannot be destroyed unless Weyl nodes with opposite chirality are annihilated with each other at the same
${\mathbf k}$ point. A Dirac node consists of a degenerate pair of coexisting Weyl nodes with opposite chirality. Although Dirac nodes
are typically unstable, for Na$_3$Bi and Cd$_3$As$_2$, they are stabilized by crystal $C_3$ and $C_4$ symmetry, respectively \cite{WANG12,LIU14,LIU14_CdAs,XU15_Dirac,WANG13_CdAs}.
The degeneracy in the Dirac nodes can be lifted when either inversion or time-reversal symmetry is broken, or both \cite{MURA07}.
Recent experiments on Na$_3$Bi, Cd$_3$As$_2$, and the WSM TaAs family showed interesting chiral magnetic effects as a result of the chiral
anomaly \cite{XION15,HUANG15_TaAs,LI16_CdAs} as well as unusual quantum oscillations in magneto-transport \cite{MOLL16}.


Topological materials are also predicted to induce new phenomena when they are in contact with conventional or non-topological materials
\cite{FU08_Majorana,HASAN10,QI11,BURK11,KATMIS16,Chen_Franz_16}. Such topological heterostructures have been widely explored in the context of
topological insulators where novel features such as Majorana modes or topologically enhanced magnetism can arise
\cite{FU08_Majorana,HASAN10,QI11,KATMIS16}, but little is known about the properties of heterostructures that include DSMs or WSMs \cite{Chen_Franz_16}.
Strong coupling between surface and bulk states in DSMs and WSMs allows interfaces to play a crucial role in hybrid structures, although the topological characteristics of DSMs and WSMs arise from the bulk. The interface may induce proximity effects or some charge transfer and
can change some orbitals near $E_F$ or break some inherent symmetries. Therefore, it may be used as a tool to probe and control their
topological bulk and surface properties. In order to investigate the coupling and interface effects, it is important to simulate hybrid
structures including the interface at the atomistic level as well as to utilize low-energy effective models simultaneously. However,
a majority of studies of DSMs and WSMs have, so far, focused on bulk properties, i.e. chiral magnetic and anomalous Hall effects
\cite{ZYUZ12,XION15,HUANG15_TaAs,LI16_CdAs,SON13,CHEN13_Axion}, and properties of the Fermi-arc surface states in the absence of the interface
solely from an effective model approach \cite{WANG12,WANG13_CdAs,NARA14,HUAN15,SOLU15,RUAN16}. Even without the interface it is extremely
challenging to investigate properties of the Dirac node or Weyl node projections and topological Fermi-arc states through direct
simulations of finite-sized DSMs and WSMs because the band gap closes very slowly with sample thickness \cite{NARA14}.
It has been known that the sizeable
band gap in DSM and WSM slabs makes it impossible to determine the positions of the Dirac node or Weyl node projections.





Here we show that even for thin DSM films with a sizeable band gap, asymmetric charge transfer at the surface allows one to accurately
identify the locations of the 3D Dirac node projections and to examine and modify properties of Fermi-arc states connecting them.
While our approach works for any DSM, we demonstrate it explicitly by simulating Na$_3$Bi films with K or Na adsorption within DFT in
conjunction with an effective model Hamiltonian. Our study functions as a first model to mimic the interface without other complications
beyond the charge transfer and ensuing effects in DSM-based hybrid structures. We consider Na$_3$Bi (1$\bar{2}$0) films with thicknesses in
the range 5.5-21.8~nm with band gaps in the range 26-75~meV. Upon adsorption, charge transfer occurs close to the top surface only,
so bulk properties such as the locations of the 3D Dirac nodes remain intact. Therefore the positions of the projections of the Dirac nodes
do not change, whereas the topological top-surface states (TSS) are shifted downward in energy relative to the bottom-surface states (BSS)
near $E_F$, due to the charge transfer. The structure of these robust surface states can thus be engineered by controlling the amount
of charge transfer. This opens up many intriguing possibilities for controlling novel surface transport phenomena such as the so called
``Weyl orbit" \cite{POTT14,MOLL16}. In addition, the shift of the TSS enables the identification of Dirac node projections despite
a band gap. Since the node locations do not change, the downward shift of the TSS leads to easily identifiable gapless crossings of the
states at the node projections. We find that the positions of the node projections are not sensitive to the amount of charge transfer,
slab thickness, or the symmetry of a non-topological adlayer or substrate surface in the hybrid structure, as long as the slab is thick
enough such that the asymmetric charge transfer does not occur throughout the slab. Furthermore, we show that there exist unique spin
textures (distinct from a topological insulator) near the Dirac node projections and that the spin textures and the separation
between the top- and bottom-surface Fermi arcs
can be modified by varying the amount of charge transfer, independently of slab thickness or the symmetry of non-DSM layers in the
heterostructures. Our findings can be observed by top or bottom gating without adatoms or interfaces.
\\
\\
{\noindent {\bf Results}}
\\

{\noindent {\bf Electronic structures of bulk and pristine slabs}}
\\

\begin{figure*}[h]
\begin{center}
\includegraphics[width=\textwidth,height=0.95\textheight,keepaspectratio]{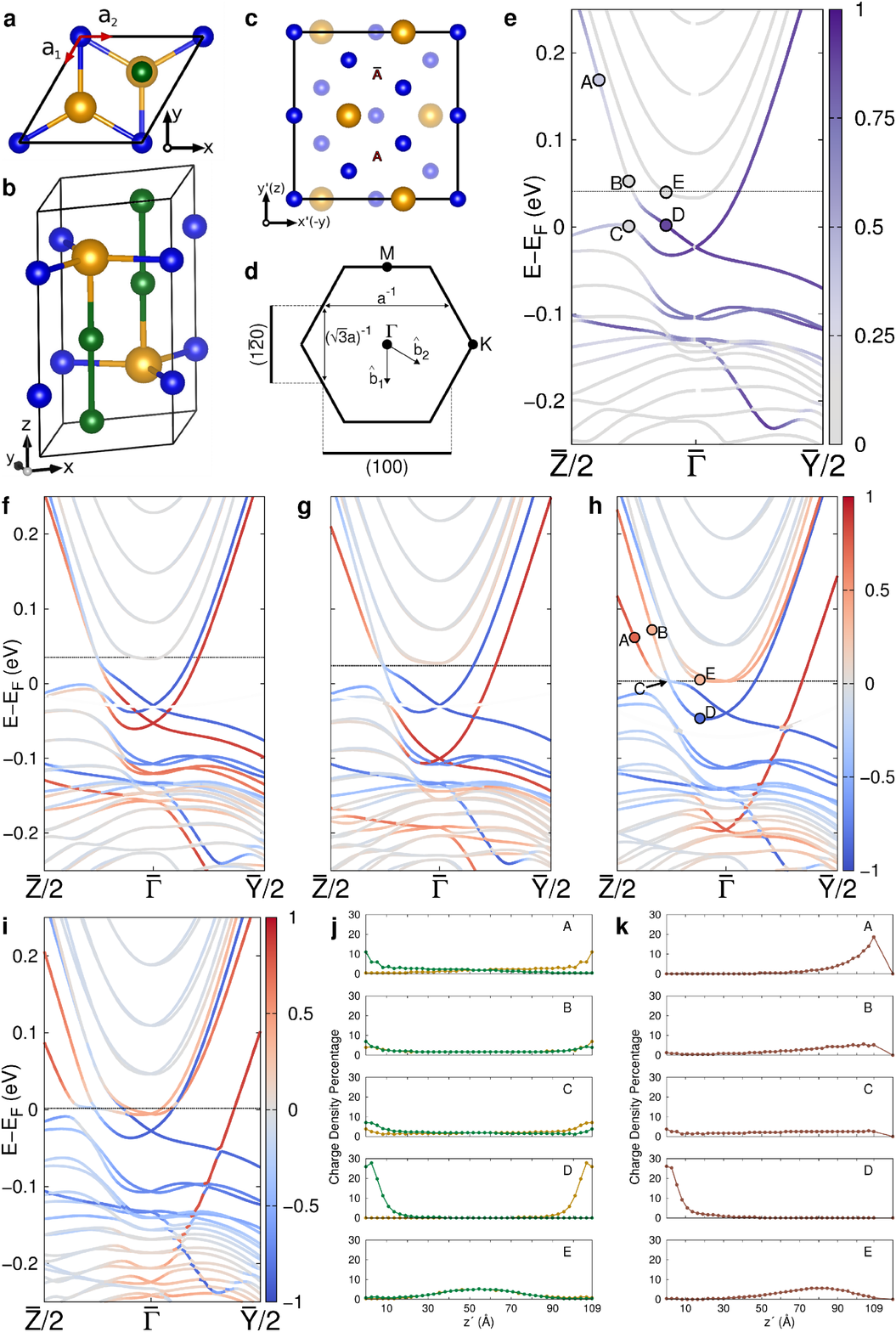}
\end{center}
\label{fig:geo}
\begin{tabular}{c}
    FIG. 1: (Continued on following page.)
\end{tabular}
\end{figure*}
\begin{figure}
\caption{{\bf Bulk and slab crystal structures of Na$_3$Bi and DFT-calculated electronic structures of (1${\bar 2}$0) slabs.} ({\bf a}) Top-down view of the bulk unit cell of Na$_3$Bi which crystallizes in the P6$_{3}/$mmc hexagonal structure. This exhibits the unit cell of the (001) surface. ({\bf b}) The bulk unit cell at a slight perspective angle. Bi atoms are large and orange. Na atoms are small and blue if in the same plane as Bi or small and green if not. The in-plane lattice vectors are shown in {\bf a}. ({\bf c}) The unit cell of the $(1\bar{2}0)$ surface. The primed (unprimed) coordinates are the local (crystal) axes. Two inequivalent atomic layers are shown, the lower layer appearing faded. ({\bf d}) A top-down view of the bulk crystal first Brillouin zone (BZ) and the projections onto two surface BZs. ({\bf e})-({\bf i}) The surface band structure of the pristine 10.9 nm slab, Case I, Case II, Case III, and Case IV as explained in the text, respectively (Na- or K-impurity bands not shown). $E_c$ is marked by a horizontal line in each case. On the color scales, absolute values close to unity indicate high surface localization of the states and are further discussed in Supplementary Note 1. ({\bf j}) Charge density vs coordinate $z'$ profiles for states identified in {\bf e}. ({\bf k}) Charge density vs coordinates $z'$ profiles for states identified in {\bf h}. }
\end{figure}


We consider bulk Na$_3$Bi with space group P6$_3$/mmc, as shown in Fig.~1a,b, with experimental lattice constants ($a=5.448$ and $c=9.655$~\AA) \cite{WANG12}. Our bulk band structure and 3D Dirac nodes ${\mathbf k_D}$ calculated by applying DFT with spin-orbit coupling
(SOC) agree well with those in the literature \cite{WANG12,LIU14,XU15_Dirac}, where ${\pm \mathbf k_D}=(0, 0, \pm 0.13)$~$2\pi/c$.
See Supplementary Fig.~1. Although some recent studies suggest that Na$_3$Bi with space group P$\bar{3}$c1 (very similar to P6$_3$/mmc) may have
a slightly lower energy than that with P6$_3$/mmc \cite{CHEN2014,JENK2016}, we stick to the space group P6$_3$/mmc in this work.
The reasons are that most of the experimental works on Na$_3$Bi \cite{WANG12,LIU14,XU15_Dirac,XION15,ZHANG14NaBi,XU15PRB}
take the crystal structure to be
P6$_3$/mmc and that Na$_3$Bi with P$\bar{3}$c1 has the 3D Dirac nodes at the same ${\mathbf k_D}$ points with the same Fermi-arc surface states \cite{CHEN2014}. The Fermi-arc surface states exist at surfaces which do not have the $c$ axis as their normal vector.
On a (001) surface the Weyl nodes with opposite chirality are projected onto the same ${\mathbf k}$ point, so Fermi-arcs are not guaranteed.
Keeping this in mind, we simulate two representative surfaces: (1$\bar{2}$0) and (100) (Fig.~1c,d). Normal vectors to the (1$\bar{2}$0) and
(100) surfaces correspond to the [0${\bar 1}$0] and [210] directions in real space. See Supplementary Fig.~2. The (1${\bar 2}$0) surface
is neutral in charge since both Na and Bi atoms appear stoichiometrically. However, the (100)
surface is polar and so dangling bond states are expected. We simulate (100) slabs with thicknesses of 5.50 and 9.44~nm with different
types of surface terminations by using DFT with SOC. Our calculated band structures show several dangling bond states near $E_F$ without
surface state Dirac cones at $\overline{\Gamma}$ (Supplementary Fig.~3). Hence, hereafter, we consider the (1$\bar{2}$0) surface only.

We simulate four pristine Na$_3$Bi (1${\bar 2}$0) slabs with thicknesses of 5.5, 10.9, 16.3, and 21.8 nm by applying DFT with SOC. The (1${\bar 2}$0)
surface has $C_2$ symmetry about the $x$ axis and mirror symmetry across the $y$ axis, as shown in Fig.~1a-c, where the unprimed
(primed) coordinates are in terms of crystal (local) axes. The calculated band structure of the 10.9 nm slab is shown in Fig.~1e, while those
of the three other thickness slabs are in Supplementary Fig.~4. Each band has double degeneracy due to inversion and time-reversal symmetry except at time-reversal invariant $k$ points (i.e., $\overline{\Gamma}$, $\overline{Z}$, and $\overline{Y}$) with fourfold degeneracy, where $k$ is two-dimensional momentum. No dangling bond states are found near $E_F$. Along the $\overline{\Gamma}-\overline{Z}$ direction, there exists
a bulk band gap. See the density of states in Supplementary Fig.~5a. Due to the finite sizes of the slabs or quantum confinement, the conduction
and valence bands do not meet. The band gap of the (1${\bar 2}$0) slab decreases very slowly with increasing slab thickness
(Supplementary Fig.~5b) in agreement with the literature \cite{NARA14}. Dirac dispersion is found near $\overline{\Gamma}$ in the vicinity of
$E_F$. We identify surface states with a given color scale by computing electron density vs vertical coordinate $z^{\prime}$, or $x$, for each band
at different $k$ points. On the color scale, absolute values close to unity indicate states which are highly localized at the surface. The
detail of the color scale is discussed in Supplementary Note 1. The electron density vs $z^{\prime}$ plots for states A-E
for the 10.9-nm slab are shown in Fig.~1j. State A has some contributions from the surface with a long non-zero tail deeply penetrating
into the slab, while the states B and C near $k_z^D$ show flat contributions from the whole slab, where $k_z^D=0.13 \cdot 2\pi/c$.
The horizontal line in Fig.~1e crosses the bands at $k_z^D$. The electron density of state D is clearly localized onto the top or bottom
surface, with a decay length of about 2~nm, whereas state E has electron density localized mainly in the middle of the slab. We assign
the states B and C as bulk states and state D as a surface state. State E is a quantum-well state. This identification is consistent
with the fact that the bulk conduction and valence bands meet at the 3D Dirac nodes and that topological surface states are expected near
$\overline{\Gamma}$ due to the large bulk band gap with band inversion at $\Gamma$ \cite{WANG12,Kane,Gorbar2015,KARG16}. See Supplementary
Fig.~1. The finite
band gap at $\Gamma$ allows one to determine the Z$_2$ invariant despite the metallic nature of Na$_3$Bi.
As shown later, existence of the surface Dirac cones near $\overline{\Gamma}$ for the adsorbed Na$_3$Bi slabs reinforces
the topological robustness of these two-dimensional Dirac cones.
Along the $\overline{\Gamma}-\overline{Z}$ direction, the surface states {\it gradually} become bulk states as
$k_z^D$ is approached from $\overline{\Gamma}$, but that is not the case along the $\overline{\Gamma}-\overline{Y}$ direction.
\\
\\
{\noindent {\bf Electronic structures of slabs with asymmetric charge transfer}}
\\

To introduce the asymmetric charge transfer, we simulate four cases of adsorbates: Na atom  at 15~\AA,~Na atom at 10~\AA,~K atom at 10~\AA,~and
two K atoms at 6~\AA~above the 10.9~nm thick slab. These are henceforth referred to as Case I, II, III, and IV, respectively. The cases of
thinner and thicker adsorbed slabs will be addressed in Discussion. Even though
a Na or K atom is adsorbed on site ``A" indicated in Fig.~1c, we confirm that the calculated electronic structure does not depend on the
adsorption site because of the large distance between the adatom and the surface. (For Case IV, one K atom is on the site ``A" and the other
on site ``$\bar{\text{A}}$.") Thus, the adatom-enriched surface maintains the same symmetry as the pristine surface.
By varying the distance between the adatom and the surface or changing the adatom type, we can vary the amount of charge transfer to
the slab. Figure~1f-i shows our DFT-calculated band structures with top and bottom surface states identified as $+1$ and $-1$
for Cases I, II, III, and IV, respectively. The surface states are determined by using the same scheme as for the pristine slabs. See
Supplementary Note 1 for further details.


For Case I, the lifting of the double degeneracy is most apparent for the surface states (except at the time-reversal invariant $k$ points,
where the quadruple degeneracy is lifted to double degeneracy) because the charge transfer occurs mainly at the top surface rather than
throughout the slab. The top surface states are shifted downward in energy relative to the bottom surface states near $E_F$, whereas the
bottom surface states have approximately the same energy as those for the pristine slab relative to $E_F$ (see $E_{\Gamma}^{BSS}$ in
Table I). At $\overline{\Gamma}$ the top-surface Dirac point is separated from the bottom-surface Dirac point by 23.6~meV (Fig.~1f). This
energy difference measures a potential difference between the top and bottom surfaces. A similar shift of the two surface Dirac cones
was shown for topological insulators with adatoms \cite{PARK13NJP}. We find that 0.027~$e$ per unit cell area
(or 0.0003~$e \cdot \AA^{-2}$)~is transferred
from the adatom to the slab for Case I. The calculation method and the depth of the charge transfer are shown in Supplementary Note 2 and
Supplementary Fig.~6. Once charge moves off the adatoms, the top surface states have lower energy because of the positive voltage created
by the positively charged adatoms. The transferred charge fills the top-surface Dirac cone only, which explains the shift of the two
surface Dirac cones near $\overline{\Gamma}$. The same effect can be reproduced by applying a gate voltage to the top surface without
either adatoms or surface doping. Interestingly, along the $\overline{\Gamma}-\overline{Z}$ direction,
two outermost conduction bands meet at $k_z^D$ with energy $E_c$ as indicated by the dashed horizontal line.
The crossing of the bands at $\pm k_z^D$ is gapless. The conduction and valence bands near $\pm k_z^D$ still have bulk characteristics,
similar to the case of the pristine slab.

For Case II, the trend is similar to Case I with an increased potential difference between the top- and bottom-surface Dirac cones.
The separation between the top- and bottom-surface Dirac points is 71.1~meV at $\overline{\Gamma}$ (Fig.~1g). The calculated
amount of the charge transfer is about 0.044~$e$ per unit cell area (or 0.0005~$e \cdot \AA^{-2}$). The transferred charge penetrates
into the slab up to about 3-4 nm (Supplementary Fig. 6). In this case, the two outermost
conduction bands also meet with each other at $\pm k_z^D$.

For Case III, the top-surface Dirac cone is buried below the bulk valence bands and coupled to bulk states even near $\overline{\Gamma}$.
The top-surface Dirac cone is found at 156.3 meV below the bottom-surface Dirac cone (Fig.~1h). The charge transfer amount is about 0.073~$e$
per unit cell area (or 0.0008~$e \cdot \AA^{-2}$). The transferred charge penetrates into the slab up to about 4~nm.
The electron density vs $z^{\prime}$ plots show surface character for state D and bulk characters for states B and C in Fig.~1k.
The electron density of state E in Fig.~1k compared to that of state E in Fig.~1j clearly reveals small downward band bending at the top
surface due to the charge transfer, similarly to the case of a topological insulator with adsorbates \cite{PARK13NJP}. For Case III
the crossings of the bands still occur at $\pm k_z^D$ (Fig.~1h).

In Case IV, the potential difference is 196.6~meV and the charge transfer amount is about 0.116~$e$ per unit cell area
(0.0013~$e \cdot \AA^{-2}$). The transferred charge penetrates into the slab up to about 4~nm.
The lowest quantum-well states right above $E_F$ exhibit larger band bending than for Case III.
For Case IV the band crossing also occurs at $\pm k_z^D$ (see Table I).
A detailed analysis of these crossing points for Cases I-IV will be discussed later.
\\

{\noindent {\bf Constant energy contours and spin texture}}
\\



\begin{figure*}
\begin{center}
\includegraphics[width=1 \textwidth, angle=0]{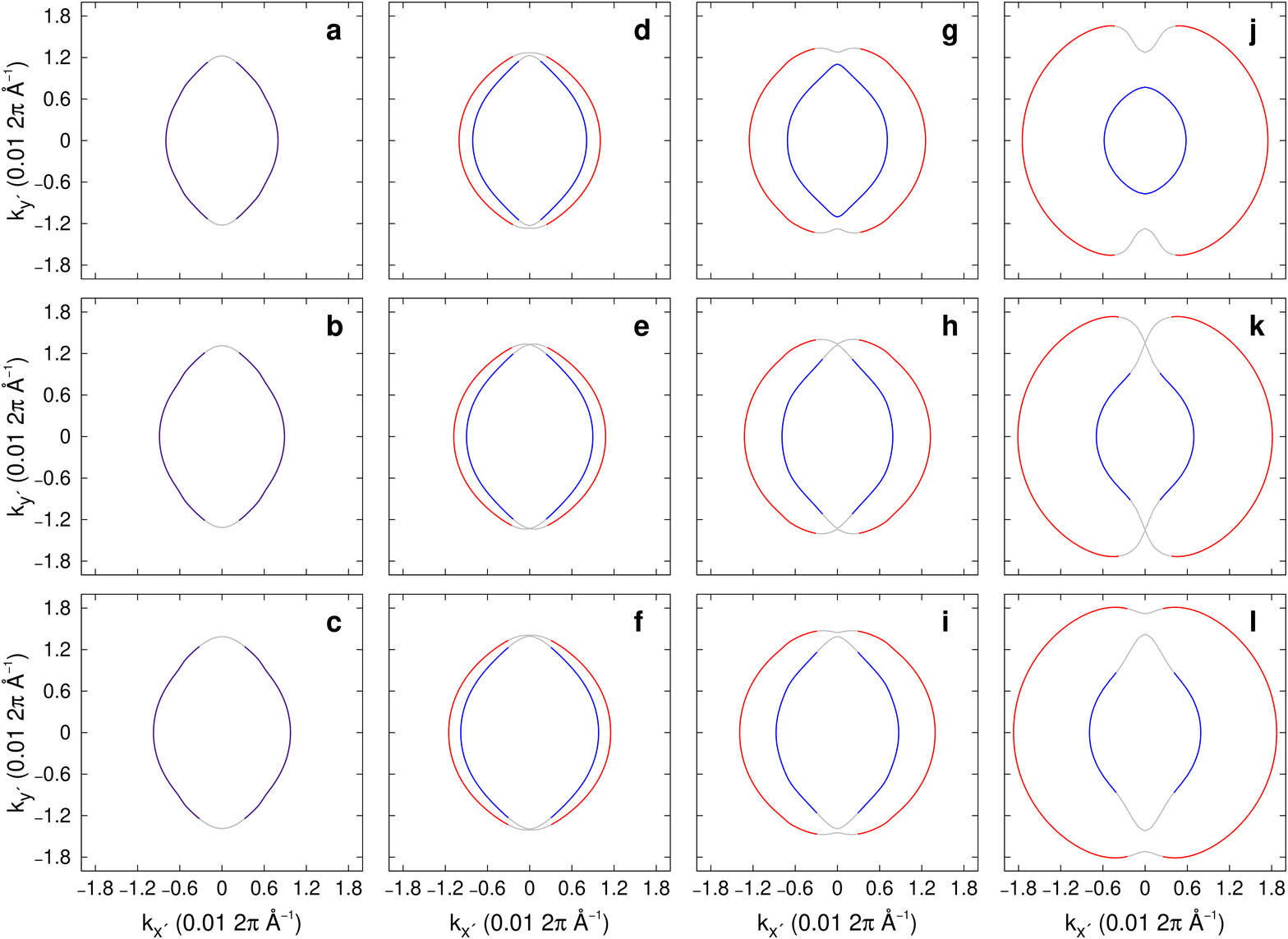}
\caption{{\bf Constant energy contours (CECs) of four systems at three different energies.} Top (bottom) surface states are identified in red (blue). The doubly degenerate surface states in the pristine case are shown in purple. The top row is at $E=E_{c}-10$ meV. The middle row is at $E=E_{c}$.
The bottom row is at $E=E_{c}+10$ meV.  The first column ({\bf a-c}) is for the pristine 10.9 nm slab. The second column ({\bf d-f}) is for Case I.
The third column ({\bf g-i}) is for Case II. The fourth column ({\bf j-l}) is for Case III. In all of the adsorbate cases, the CECs only touch each other at $E=E_{c}$.}
\label{fig:CEC}
\end{center}
\end{figure*}

\begin{figure*}
\begin{center}
\includegraphics[width=1 \textwidth, angle=0]{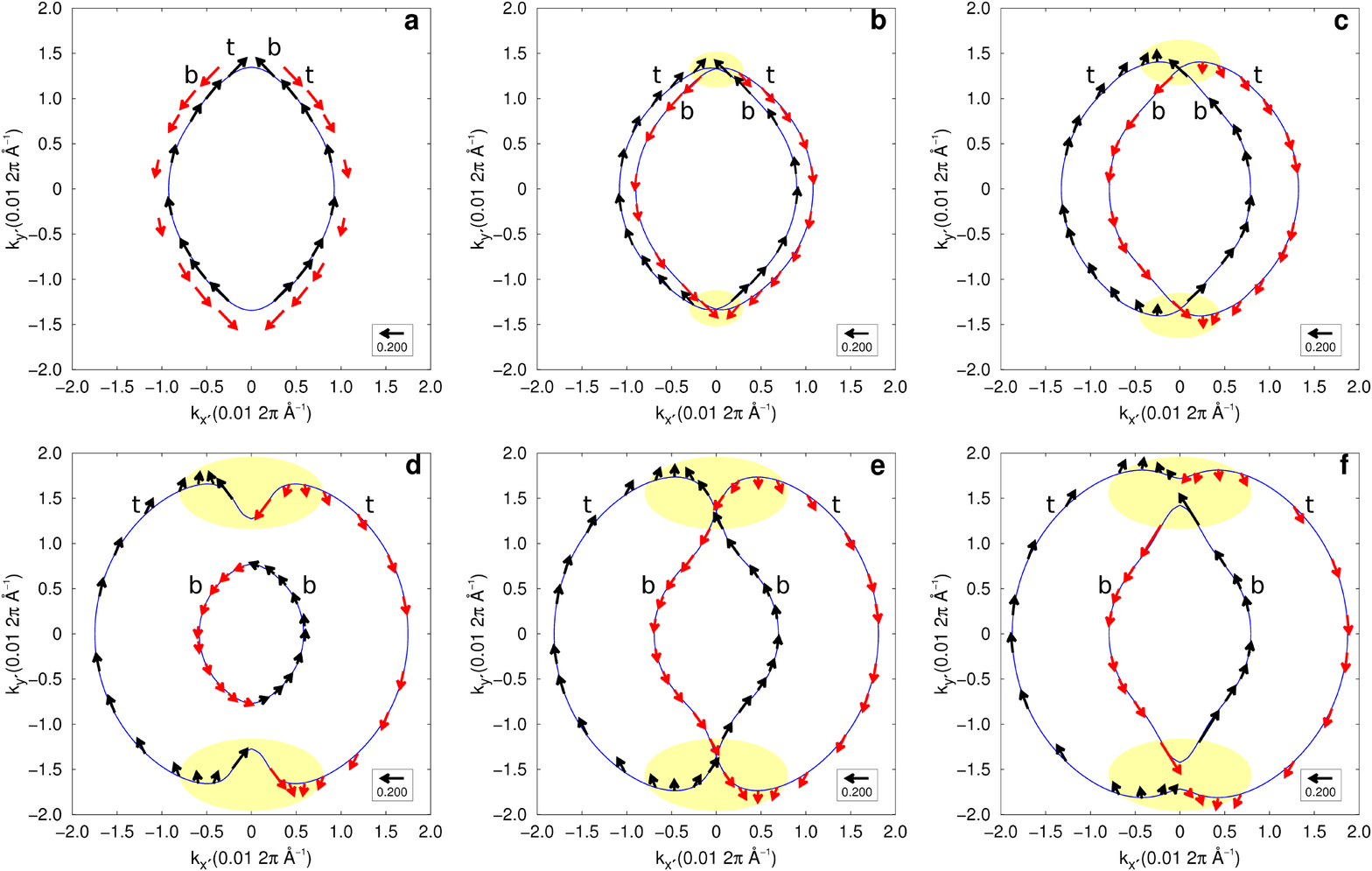}
\caption{{\bf Spin textures along constant-energy contours.} ({\bf a}) For the pristine slab at $E=E_{c}$. The degenerate bands have exactly opposite spin, and the spin texture for the down-sector is shown offset from the CEC for clarity. ({\bf b}) For Case I at $E=E_{c}$. ({\bf c}) For Case II at $E=E_{c}$. ({\bf d}) For Case III at $E=E_{c}-10$ meV. ({\bf e}) For Case III at $E=E_{c}$. ({\bf f}) For Case III at $E=E_{c}+10$ meV. In every panel, the spin is colored according to which spin-sector it belongs to, up (down) -sector spin is shown in black (red). The top (t) or bottom (b) surface character is also marked. In the adsorbate cases, the region including $k_z^D$ and ($\pm k_{x^{\prime}}^0$,$\pm k_{y^{\prime}}^0$) is
highlighted in yellow.}
\label{fig:spin}
\end{center}
\end{figure*}

For a better understanding of the band crossing points and the effect of the charge transfer, we examine constant energy contours (CECs) of the
two outermost conduction bands near $E_c$ and their spin textures at $E_c$ for the pristine 10.9-nm slab. For the pristine slab we select the
$E_c$ value to be the energy of the conduction bands at $k_z^D$, indicated with a horizontal line in Fig.~1e.
Figure~\ref{fig:CEC}a-c shows DFT-calculated CECs at $E_c-10$~meV, $E_c$, and $E_c+10$~meV, where colored bands are surface states. The doubly degenerate contours are symmetric about the local $k_{x^{\prime}}$ (crystal $k_y$) and $k_{y^{\prime}}$ ($k_z$) axes. The contours at $E_c$
belonging to a single surface intersect the $k_{y^{\prime}}$ axis at 90$^{\circ}$ degrees (without singularity) rather than obliquely (with singularity) due to the quantum confinement (Fig.~\ref{fig:CEC}b). Near the $k_{y^{\prime}}$ axis the bands have bulk characteristics.
Figure~\ref{fig:spin}a illustrates DFT-calculated spin textures of the bands at $E_c$. The spin polarization of the bands is tangential to the
CEC. Along the $k_{y^{\prime}}$ ($k_{x^{\prime}}$) axis, the spin polarization is constrained to be normal to the $k_{y^{\prime}}$
($k_{x^{\prime}}$) axis because of the glide mirror (horizontal mirror) plane $k_{y^{\prime}}$-$k_{z^{\prime}}$
($k_{x^{\prime}}$-$k_{z^{\prime}}$). The top-surface (bottom-surface) states show clockwise (counter-clockwise) rotation of the spin, which
will be re-interpreted in comparison to the spin textures for the adsorbed slabs later.

Next we investigate CECs of the two conduction bands for Cases I-IV near $E_c$. Figure~\ref{fig:CEC}d-l shows the CECs for Cases I-III at
$E_c-10$~meV, $E_c$, and $E_c+10$~meV with band characteristics (i.e., surface or bulk states). The CECs for Case IV are not shown since
they have qualitatively similar features to those for Case III. In all four cases, the contours are still symmetric about the
$k_{x^{\prime}}$ and $k_{y^{\prime}}$ axes, and they cross at $(k_{x^{\prime}},k_{y^{\prime}})$$=$$(0, \pm k_z^D)$ with {\it singularity}
only for energy $E_c$ (see Table I). The contours for $E_c$ connecting the $(0, \pm k_z^D)$ points have bulk character near the crossing
points and gradually change to top or bottom surface-state character away from the points (Fig.~\ref{fig:CEC}e,h,k). These features of the
$E_c$ contours coincide with those of the Fermi arcs appearing on a surface of a DSM when the $(0, \pm k_z^D)$ points are projections of the
Dirac nodes. Our analysis (based on an effective model) that will be discussed later ensures that these contours are indeed Fermi arcs.
Away from $E_c$ the closed contours do not cross each other. Although near the crossing points,
$\pm k_z^D$, the surface state character disappears for both contours at a given energy, the contours are referred to as top- and
bottom-surface contours for convenience in explaining their features and spin textures. In Cases I-IV, the bottom-surface contour always
appears inside the top-surface contour. For the small charge transfer (Case I), the overall contour shapes do not change much with
energy except for the region very close to the crossing points (Fig.~\ref{fig:CEC}d-f). As the amount of charge transfer increases, the
top-surface contour at $E_c$ expands along both the $k_{x^{\prime}}$ and $k_{y^{\prime}}$ directions with a slightly pinched circular
shape, whereas the bottom-surface contour shrinks mostly along the $k_{x^{\prime}}$ axis while keeping its oval or almond shape.
Compare Fig.~\ref{fig:CEC}e with Fig.~\ref{fig:CEC}h,k. For large charge transfer (Cases III and IV), the overall shapes of the CECs now
significantly change at energies away from $E_c$. Compare Fig.~\ref{fig:CEC}k with Fig.~\ref{fig:CEC}j,l. Below $E_c$ the bottom-surface
closed contour has pure surface-state character without singularity for all $k$ values (Cases II-IV).


We also examine spin textures at the $E_c$ contours for Cases I-IV. Figure~\ref{fig:spin}b,c,e shows DFT-calculated spin textures of the
states at the $E_c$ contours for Cases I-III, respectively. For all four cases the spin polarization along the $k_{y^{\prime}}$ axis is
constrained to be normal to the mirror $k_{y^{\prime}}$-$k_{z^{\prime}}$ plane, but at the crossing points, $\pm k_z^D$, the spin is
ill-defined. For small charge transfer (Case I), the spin polarization along the bottom-surface contour rotates
counter-clockwise and is tangential to the contour except for very close to the $k_{y^{\prime}}$ axis. The spin orientations along the top-surface contour are opposite to those along the bottom-surface contour. Overall features of the spin textures
in this case are similar to those for the pristine slab. In Case II, interestingly, the spin texture along the top-surface
contour is found to qualitatively differ from that for Case I and the pristine slab, as highlighted in the yellow shaded areas in
Fig.~\ref{fig:spin}c, although the spin texture along the bottom-surface contour is similar to that for Case I. Near the
crossing points, the spin polarization along the top-surface contour is substantially reduced and its orientation deviates
considerably from the tangent to the contour. Approaching $k_z^D$, the spin polarization in the $k_{x^{\prime}} > 0$ plane ($k_y < 0$ plane)
becomes almost perpendicular to the contour and then it turns toward the other direction such that it smoothly interpolates into the
spin polarization of the bottom-surface contour in the $k_{x^{\prime}} < 0$ plane. The same trend is found in Case III
with more pronounced features and it appears at larger $|k_{x^{\prime}}|$ and $|k_{y^{\prime}}|$ values due to the larger charge transfer.
The same trend as Case III is also observed for the spin textures for Case IV (not shown). The ($\pm k_{x^{\prime}}^0$,$\pm k_{y^{\prime}}^0$)
points at which the spin polarization of the states at the top-surface contour is almost normal to the contour for Cases II-IV are listed
in Table I. This unique spin texture of the top-surface contour is distinct from that associated with the surface states of a topological
insulator. The interesting spin texture still appears at slightly below and above $E_c$, as shown in Fig.~\ref{fig:spin}d,f.
\\
\\

{\noindent {\bf Analysis based on DFT calculations and model}}
\\

\begin{figure*}
\begin{center}
\includegraphics[width=1 \textwidth, angle=0]{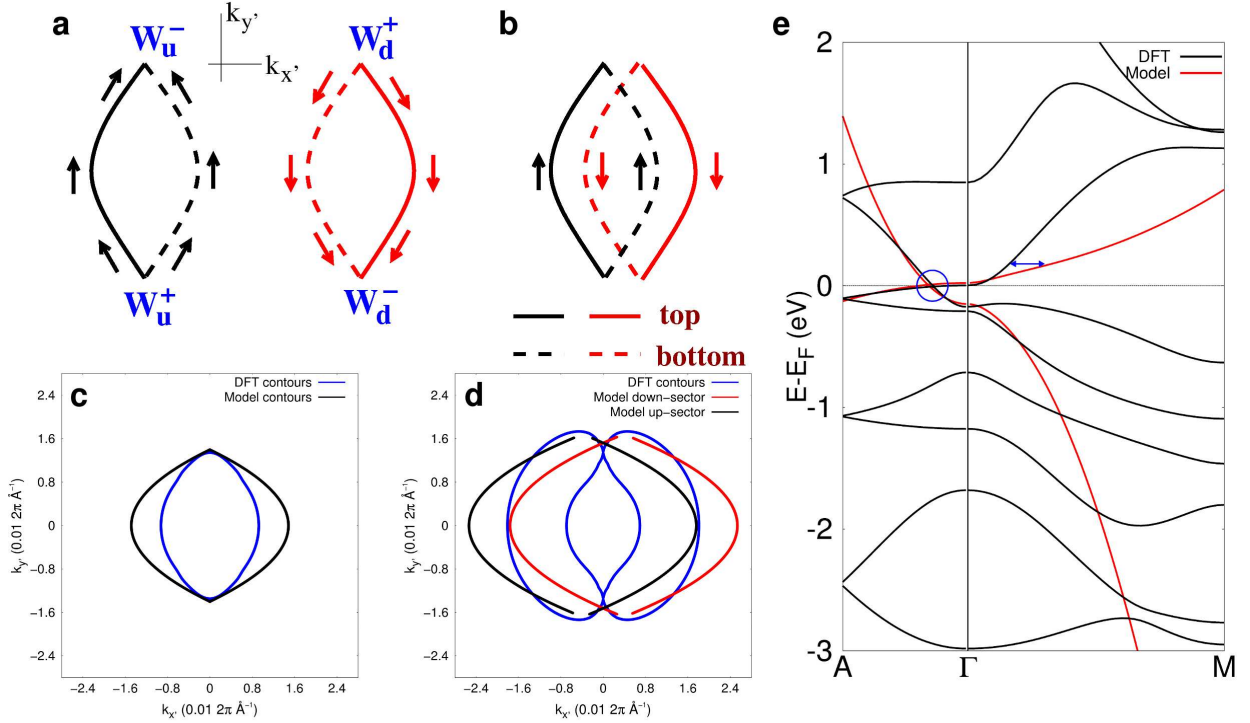}
\caption{{\bf Comparison of the DFT calculation with the effective model.} ({\bf a}) Schematic diagram of the Fermi-arc surface states connecting
Weyl node projections from each spin-sector for the up-sector (black) and down-sector (red). The top (bottom) surface character is shown by the solid (dashed) curves. In the pristine slab case, these contours lie on top of each other, and so the schematic in this panel is not meant to indicate absolute locations of the Weyl node projections in the surface BZ.
({\bf b}) Schematic diagram of the surface states in the adsorbate cases, with spin and surface character denoted as in {\bf a}. ({\bf c}) CECs
from the model for the pristine semi-infinite slab and from the DFT calculation for the pristine 10.9 nm slab. ({\bf d}) CECs from the model with $V=0.12$~eV and from the DFT calculation for Case III. ({\bf e}) DFT- and model-calculated bulk band structure. The model locates the bulk Dirac node slightly further from $\Gamma$ than our DFT calculation does, circled in blue. The blue arrow draws attention to the significant deviation of the bands of the model from our DFT calculation along the ${\Gamma-\text{M}}$ direction, which is responsible for the substantial difference between the DFT and model CECs along the local $x^{\prime}$ axis, as seen in {\bf c} and {\bf d}.}
\label{fig:model}
\end{center}
\end{figure*}

Now we analyze our DFT-calculated results by using an effective model. With inversion symmetry, the electronic structure of bulk Na$_3$Bi
near $\Gamma$ can be described by an effective low-energy $4 \times 4$ model Hamiltonian \cite{WANG12} (see Methods). The same Hamiltonian
with different parameter values can be used for the electronic structure of bulk Cd$_3$As$_2$ \cite{WANG13_CdAs}. The effective model up to
second order in ${\mathbf k}$ indicates that Weyl fermions associated with one pair of Weyl nodes, $W_u^{\pm}$, are related to Weyl
fermions at another pair of Weyl nodes, $W_d^{\mp}$, by interchanging $k_y$ with $-k_y$ \cite{Gorbar2015}. Thus, the Fermi-arc
surface states connecting projections of $W_u^{\pm}$ are also related to those connecting projections of $W_d^{\mp}$ by
$k_y \leftrightarrow -k_y$ or $k_{x^{\prime}} \leftrightarrow -k_{x^{\prime}}$ (Fig.~\ref{fig:model}a). The two pairs of Weyl nodes
($W_u^{\pm}$, $W_d^{\mp}$) are referred to as {\it up}- and {\it down}-sector following the (crystal) $z$ component of the spin polarization of
the Fermi-arc states \cite{Gorbar2015}. The calculation based on the model \cite{Gorbar2015} reveals that
for the up-sector WSM, the Fermi-arc state from the top surface appears in the $k_{x^{\prime}} < 0$ ($k_y > 0$) plane, while that from the
bottom surface appears in the $k_{x^{\prime}} > 0$ ($k_y < 0$) plane. For the down-sector WSM, the Fermi-arc states from the top and bottom
surfaces are formed on half-planes opposite to those for the up-sector (Fig.~\ref{fig:model}a).
These spin textures qualitatively agree with our DFT-calculated results for the pristine slab (Fig.~\ref{fig:spin}a vs \ref{fig:model}a).
The model-calculated CECs at $E_c$ are compared to the DFT-calculated CECs in Fig.~\ref{fig:model}c.
The $|k_z^D|$ value from the model, $\sqrt{M_0/M_1}$ ($=0.01438 \cdot 2\pi$~\AA$^{-1}=0.14 \cdot 2\pi/c$), is slightly greater than that
from our DFT band structure and the literature \cite{WANG12}. See Methods for details. The Fermi-arc states
from the model meet the $x^{\prime}$ axis at a much greater $k_{x^{\prime}}$ value than that from the DFT result
(Fig.~\ref{fig:model}c). This discrepancy
is not due to the small slab thickness but arises from differences in the bulk band structure obtained from the model and DFT
calculations (of ours and the literature \cite{WANG12}), as shown in Fig.~\ref{fig:model}e.
The parameters used in the effective model give rise to dispersion in agreement with the DFT-calculated bands along the $z$ axis with
a slightly larger value of $|k_z^D|$ than 0.13~$\cdot 2\pi/c$, whereas the dispersion along the $y$ axis is greatly off from the
DFT-calculated bands except for in the vicinity of $\Gamma$.

Now with the asymmetric charge transfer at the {\it surface}, the symmetry in the bulk does not change and so the locations of the 3D Dirac
nodes arising from the bulk are not affected, i.e. $\pm$${\mathbf k_D}$$=$(0, 0, $\pm$0.13~$\cdot 2\pi/c$). Thus, the Fermi-arc states
from the top and bottom surfaces must connect the projections of the Dirac nodes. This is because of the coexistence of the degenerate
Weyl nodes of opposite chirality at the Dirac nodes; owing to the up/down sector symmetry \cite{Gorbar2015}, the degenerate Weyl nodes
(W$_u^{-}$ and W$_d^{+}$ in Fig.~\ref{fig:model}a) are not annihilated at each Dirac node but instead are superimposed on each other.
The Weyl node with which a given Weyl node would annihilate is actually situated across the Brillouin zone at the opposite Dirac node.
Our DFT calculations show such connections between the Dirac node projections. Compare Fig.~\ref{fig:CEC}e,h,k and
Fig.~\ref{fig:spin}b,c,e with Fig.~\ref{fig:model}b. These connections explain the existence of the gapless band crossings at $\pm k_z^D$
and why the bands shown in Fig.~\ref{fig:CEC}e,h,k carry bulk characteristics near the Dirac node projections. The shapes of
the top- and bottom-surface contours can be explained by introducing a surface potential $V$ to the top-surface states in the model since the
charge transfer occurs mainly at the top surface. See Methods for details. In the semi-infinite slab limit, with $V=0.12$~eV, we calculate
the top- and bottom-surface states at $E_F$ for the up- and down-sector within the model, as shown in Fig.~\ref{fig:model}d. The value of $V$ qualitatively corresponds to Case III. The model-calculated CECs of the surface states show that the down-sector surface states are shifted
toward the positive $k_{x^{\prime}}$ direction, while the up-sector surface states shift toward the opposite direction,
with their crossing points near $\pm k_z^D$. This trend agrees with our DFT-calculated CECs of the Fermi-arc states. With $V=0.12$~eV,
the $k$ values of the crossing points and the amount of the shift of the Fermi surface states obtained from the model (represented by
$(k_{x^{\prime}}^0, k_{y^{\prime}}^0)$) are in reasonable agreement with those of our DFT calculations (Fig.~\ref{fig:model}d).
The spin textures along the top- and bottom-surface contours from our DFT calculations also qualitatively agree with those from the
model. The gapless band crossings at
$\pm k_z^D$ can be also justified by the crossings of different spin sectors, i.e. up- and down-sector. Interestingly, although the spin
textures near $(\pm k_{x^{\prime}}^0, \pm k_{y^{\prime}}^0)$ points (the yellow shaded areas in Fig.~\ref{fig:spin}c,e) follow as if
they are the Dirac node projections (Fig.~\ref{fig:model}b), these $k$ points are not the true projections of the 3D Dirac nodes. This is because the crystal $C_3$ symmetry forbids the 3D Dirac nodes from being removed from the crystal axis, and likewise the projections remain on the $k_{y'}$ axis. Therefore, our result suggests that apparent projections of the 3D Dirac nodes such as
$(\pm k_{x^{\prime}}^0, \pm k_{y^{\prime}}^0)$ at the surface are {\it not} necessarily related to the 3D Dirac nodes in the bulk. See
Supplementary Discussion 1 for our discussion of the case where asymmetric charge transfer occurs throughout the entire slab and thus breaks
$C_3$ symmetry and inversion symmetry at the bulk level. Note that this case qualitatively differs from our asymmetric charge transfer
at the surface.
\\
\\
{\noindent {\bf Discussion}}
\\


Regarding the DFT-calculated features, five caveats need to be addressed. First, simple charge transfer to the slab is not enough to observe
the main outcomes of this work. The charge transfer must be asymmetric and only at the surface. We consider the case that one Na atom is
adsorbed on both the top and bottom surfaces of the 10.9-nm thick slab while keeping the distance between the adatom and the surface at 10~\AA.
The charge transfer amount, 0.089~$e$ per unit cell area, is comparable to that for Case III. However, the calculated band structure shows
that the bulk band gap is 27.3~meV and that there are no band crossings at $\pm k_z^D$ or any other $k$ points near the Fermi level
(Supplementary Fig.~7). Second, we find that the gapless band crossings at the 3D Dirac node projections appear independently of the amount
of charge transfer and slab thickness, as long as the slab is thick enough such that the asymmetric charge transfer occurs only at the surface
rather than throughout the entire slab. We investigate the electronic structures of adsorbed slabs thinner and thicker than the 10.9~nm slab
studied earlier in the text: slabs with thicknesses of 2.7, 5.5, and 16.3~nm. See Supplementary Fig.~8. The band structure of the K-adsorbed
2.7-nm slab (where the K atom is 10 \AA~above the surface) shows gapless band crossings at $\pm$0.12~$\cdot 2\pi/c$, which does not
coincide with the Dirac node projections. However, for the 5.5~nm and 16.3~nm thick slabs with adatoms, the gapless band crossings occur
at $\pm k_z^D$. In addition, for a given charge transfer amount, the potential difference between the top and bottom surfaces does not
change with slab thickness as long as the slab is somewhat thicker than the penetration depth of the charge transfer. With a fixed amount
of charge transfer, we find that the CECs of the two conduction bands at $E_c$ and the spin textures along these contours do not change
with slab thickness. Compare Figs.~\ref{fig:CEC}k,\ref{fig:spin}e with Supplementary Fig.~9a,b.
Third, the aforementioned CECs and spin textures depend only on the amount of charge transfer, independent of slab thickness and adatom type
when the slab is thick enough. Considering the symmetry of Cd$_3$As$_2$ (with space group I4$_1$acd \cite{MAZH14}), similar spin textures to 
our work are expected for a (112) surface of Cd$_3$As$_2$ \cite{Liang2015}. Fourth, due to robustness of the Dirac or Weyl
nodes against surface perturbations, the symmetry of a non-topological material (substrate or overlayer) at the interface with a DSM or WSM is
not important in hybrid structures in order to experimentally observe our DFT-calculated properties of the Fermi-arc states and bulk states.
What is important is the amount of asymmetric charge transfer. Fifth, the gapless band crossings would not occur in thin films for the surface
where the Weyl nodes with opposite chirality for a given sector are projected to the same ${\mathbf k}$ points, such as the (001) surface,
even with the asymmetric charge transfer.

Before concluding, we would like to mention some other works which have studied different aspects of disorder in DSMs and WSMs.  The robustness of the 3D Dirac or Weyl nodes against surface disorder was discussed in the context of a linearized
Hamiltonian \cite{POTT14} and a DFT calculation of a thick TaAs slab with K adsorption \cite{TaAs_slabDFT}. Scattering of Fermi-arc
surface states over impurities in DSMs or WSMs was investigated from observation of quasiparticle interference patterns 
\cite{JEON14,INOUE16}. Stability of the Fermi-arc surface states over bulk disorder in DSMs and stability of the Weyl nodes over 
local impurities were discussed in Refs.~[\onlinecite{KARG16,BALA13}]. However, the scope and main results of our work are distinct 
from those of Refs.[~\onlinecite{POTT14,KARG16,TaAs_slabDFT,JEON14,INOUE16,BALA13}].


In summary, we investigated probing and engineering topological bulk and surface properties of DSM thin films with a band gap by the asymmetric
charge transfer which may occur at the interfaces of heterostructures involving DSMs or WSMs. We simulated K- or Na-adsorbed
Na$_3$Bi~(1${\bar 2}$0) films where a small amount of charge is transferred at the top surface only, by using DFT including SOC rather 
than using an effective model or tight-binding model with bulk parameter values. We found that the asymmetric charge transfer 
induces the gapless band crossings exactly at the 3D Dirac node projections for thin slabs despite the band gap because the 
topological top- and bottom-surface states separated by the charge transfer are pinned by the projections of the topologically robust 
3D Dirac nodes. We also showed that the CECs and spin textures at the band crossing energy can be modified by varying the amount of 
charge transfer. The main features we discussed are independent of slab thickness when the asymmetric charge transfer occurs only in close proximity to the surface. Although we considered a specific example of Na$_3$Bi slabs, our findings can be applied to
other DSMs and WSMs. 
\\
\\

\setlength{\tabcolsep}{14pt}
\begin{table*}[!htb]
\caption{{\bf DFT-calculated parameter values for the pristine and adsorbed (1{\bf$\bar{2}$}0) slabs}: the energy of the 3D Dirac node $E_c$,
the amount of charge transfer per unit cell area (u.c.a), $\Delta\rho$, the energy of the TSS and BSS at $\Gamma$ ($E_{\Gamma}^{TSS}$ and
$E_{\Gamma}^{BSS}$), $\mu$$=$$|E_{\Gamma}^{TSS}$$-$$E_{\Gamma}^{BSS}|$, the position of the projections of the 3D Dirac node along the
$k_z$ axis ($k_z^D$), the intersections of the surface states with the $k_{x^{\prime}}$ axis ($k_{x^{\prime},TSS}$ and $k_{x^{\prime},BSS}$),
and the coordinates ($k_{x^{\prime}}^0$, $k_{y^{\prime}}^0$) of the shift of the TSS.  All energies are relative to $E_F$ and all momenta
appear in units of $2\pi~$\AA$^{-1}$.}
\begin{tabular}{c | c c c c c}
\hline\hline
& Pristine & Case I & Case II & Case III & Case IV \\ [0.5ex]
\hline
$E_{c}$ (meV) & 41 & 35 & 24 & 3.3  & 2.0\\
$\Delta\rho~(e/\text{u.c.a.})$ & 0 & 0.027 & 0.044 & 0.073 & 0.116\\
$E_{\Gamma}^{TSS}$ (eV) & -0.0232 & -0.0529 & -0.1004 & -0.1957 & -0.2242\\
$E_{\Gamma}^{BSS}$ (eV) & -0.0232 & -0.0293 & -0.0293 & -0.0394 & -0.0277\\
$\mu$ (eV)  & 0 & 0.0236 & 0.0711 & 0.1563 & 0.1965\\
$k_{z}^{D}$ & 0.01346 & 0.01334 & 0.01337 & 0.01353 & 0.01377\\
$k_{x^{\prime},TSS}$ & 0.00889 & 0.01080 & 0.01323 & 0.01810 & 0.02020\\
$k_{x^{\prime},BSS}$ & 0.00889 & 0.00898 & 0.00799 & 0.00709 & 0.00541\\
$k_{y^{\prime}}^0$ & \text{N/A} & 0.01337 & 0.01405 & 0.01735 & 0.01916 \\
$k_{x^{\prime}}^0$ & \text{N/A} & 0.00057 & 0.00218 & 0.00458 & 0.00483 \\

\hline
\end{tabular}
\end{table*}

\newpage

{\noindent {\bf Methods}}
\\
{\noindent {\bf DFT Method}}
\\


We simulate the Na$_3$Bi bulk and slabs by using DFT code VASP \cite{VASP1996a,VASP1996b}. We use the generalized gradient approximation
(GGA) \cite{Perd1996} for the exchange-correlation functional and projector-augmented wave (PAW) pseudopotentials \cite{PAW,VASP1999}. Spin-orbit
coupling is included self-consistently within the DFT calculation. We use the experimental lattice constants $a=5.448$ and $c=9.655$~\AA~
\cite{WANG12}, and an energy cutoff of 250~eV for the bulk and slabs. For the bulk an $11 \times 11 \times 11$ $k$-point mesh is used.
In addition to inversion symmetry, the bulk has four mirror symmetry planes ($xy$ and $yz$ planes, and two equivalent planes
to the $yz$ plane by $C_3$ symmetry about the $c$ axis). See Fig.~1a,b. Note that our coordinates are rotated by $\pi/2$
counter-clockwise from the coordinates used in Refs.~\onlinecite{WANG12,Gorbar2015}. For the slabs, a thick vacuum layer of 30-40~\AA~is
included in the supercells. For the (100) surface, we consider a 5.50~nm thick slab with Na termination (36 atomic layers) and a 9.44~nm
thick slab with Na and Bi termination (60 atomic layers). The surface area of each supercell in this case is $a \times c$. We use a $9 \times 9 \times 1$ or $11 \times 7 \times 1$ $k$-point mesh. For the (1${\bar 2}$0) surface, we consider four different slab thicknesses,
such as 5.5, 10.9, 16.3, and 21.8~nm, which correspond to 21, 41, 61, and 81 atomic layers, respectively. The surface area of each
supercell for this surface is $\sqrt{3}a \times c$. We use a $5 \times 5 \times 1$ $k$-point mesh. We check that our calculated band
structures do not change with dipole corrections. We confirm that relaxation of the experimental geometry/structure does not change
the main results of our work.
\\

{\noindent {\bf Model Hamiltonian}}
\\


To describe the bulk Na$_3$Bi we use the effective model Hamiltonian
${\cal H}= \epsilon_0({\mathbf k}) I_{4 \times 4} + {\tilde{\cal H}}$ constructed by using a basis set
$\{ |S_{1/2}^{+}, 1/2 \rangle$, $|P_{3/2}^{-}, 3/2 \rangle$, $|S_{1/2}^{+}, -1/2 \rangle$, and $|P_{3/2}^{-}, -3/2 \rangle \}$, where the
subscripts 1/2 and 3/2 are total angular momenta ${\mathbf J}$ of $s$ and $p$ orbitals, $\pm 1/2$ and $\pm 3/2$ are projections of
${\mathbf J}$ onto the $c$ or $z$ axis, and the superscripts $\pm$ correspond to the parity values. Here
$\epsilon_0 ({\mathbf k}) = C_0 + C_1 k_z^2 + C_2 (k_x^2 + k_y^2)$ and $I_{4 \times 4}$ is a $4 \times 4$ identity matrix.
The Hamiltonian was first suggested in Ref.~\onlinecite{WANG12}. The Hamiltonian
${\tilde{\cal H}}$ reads
\[ \left( \begin{array}{cccc}
  M(k)  & iAk_{+} &  0            &    B(k)^{\star} \\
 -iAk_- & -M(k)   &  B(k)^{\star} &     0      \\
    0   &   B(k)  &     M(k)      &  iAk_-   \\
  B(k)  &     0   &   -iAk_+      &  -M(k)
\end{array}  \right) , \]
where $M({\mathbf k})=M_0-M_1 k_z^2 - M_2 (k_x^2 + k_y^2)$, $k_{\pm}=k_x \pm ik_y$, $B({\mathbf k})=\alpha k_zk_+^2$, and
$B({\mathbf k})^{\star}$ is its complex conjugate. Here
$C_0=-0.06382$~eV, $C_1=8.7536$~eV~\AA$^2$, $C_2=-8.4008$~eV~\AA$^2$, $M_0=-0.08686$~eV, $M_1=-10.6424$~eV~\AA$^2$, $M_2=-10.3610$~eV~\AA$^2$,
and $A=2.4598$~eV~\AA, from fitting \cite{WANG12} of the bulk band structure near $\Gamma$, but the parameter $\alpha$ is unknown.
Note that our Hamiltonian ${\tilde{\cal H}}$ is slightly modified from the literature, considering that our coordinates differ from those in
Refs.~\onlinecite{WANG12,Gorbar2015}. For $\alpha=0$, the lower $2 \times 2$ matrix of ${\cal H}$ can be obtained from the
upper $2 \times 2$ matrix by replacing $k_y$ by $-k_y$, and the upper (lower) $2 \times 2$ matrix represents the up-sector (down-sector)
Weyl fermions \cite{Gorbar2015}. For simplicity, henceforth we consider $\alpha=0$ or $B({\mathbf k})=0$.
After applying a canonical transformation \cite{Gorbar2015} to the {\it down}-sector $2 \times 2$ matrix ${\cal H}^{\prime}$
with a unitary matrix ${\cal U}=(I_2 + i {\sigma_y})/\sqrt{2}$, we obtain the transformed Hamiltonian
${\cal H}^d_{2 \times 2}={\cal U}^{-1}{\cal H}^{\prime}{\cal U}$ as
\[ \left( \begin{array}{cc}
\epsilon_0({\mathbf k}) -A k_y      &  M({\mathbf k}) + i A  k_x \\
M({\mathbf k}) - i A k_x               &  \epsilon_0({\mathbf k}) + A k_y
\end{array} \right) . \]
The up-sector $2 \times 2$ Hamiltonian matrix ${\cal H}^u_{2 \times 2}$ can be obtained from ${\cal H}^d_{2 \times 2}$ by
interchanging $k_y$ with $-k_y$. From ${\cal H}^u_{2 \times 2}$ and ${\cal H}^d_{2 \times 2}$, we obtain energy eigenvalues of
$\epsilon_0({\mathbf k}) \pm \sqrt{M^2({\mathbf k}) +  A^2(k_x^2 + k_y^2)}$. Here the square root vanishes at
$(0, 0, \pm \sqrt{M_0/M_1})$, which are the 3D Dirac nodes.

We first examine the Fermi-arc surface states in the absence of the asymmetric charge transfer. We focus on the down-sector Hamiltonian,
considering a semi-infinite slab occupying $x > 0$. The top surface is located at $x=0$, while the vacuum is on the $x < 0$ side.
After replacing $k_x$ by $-i \partial_x$, ${\cal H}^d_{2 \times 2}$ becomes surface Hamiltonian ${\cal H}^d_s$:
\[ \left(\begin{array}{cc}
C_0 + C_1 k_z^2 + C_2 (-\partial^2_x + k_y^2) - A k_y       &  -M_1(k_z^2 - m) - M_2 (-\partial^2_x + k_y^2) + A \partial_x   \\
-M_1(k_z^2 - m) - M_2 (-\partial^2_x + k_y^2) - A \partial_x  &  C_0 + C_1 k_z^2 + C_2 (-\partial^2_x + k_y^2) + A k_y
\end{array} \right) , \]
where $\sqrt{m}=\sqrt{M_0/M_1}$. The top surface states at $x=0$ must have a form of $e^{-px}$ with $p > 0$.
Eigenvectors of ${\cal H}^d_s$ can be written as a linear combination of $e^{-p_1 x}(1,Q_1)$ and $e^{-p_2 x}(1,Q_2)$,
where $p_1$ and $p_2$ are positive. Thus, applying the boundary conditions at $x=0$ \cite{Shan2010,Gorbar2015}, we find that the
Fermi-arc surface states at a given energy $E$ can be obtained by solving the following equations numerically.
\begin{eqnarray}
Q_1 &=& Q_2, \: \: \: \: \: \: \:
Q_i \equiv \frac{C_0 + C_1 k_z^2 + C_2 (-p_i^2 + k_y^2) - A k_y - E}{M_1(k_z^2 - m) + M_2(-p_i^2 + k_y^2) + A p_i},
\label{eq:Q} \\
p^2_{1,2} &\equiv& \frac{1}{2(C_2^2 - M_2^2)} [ 2(C_2 K + M_2 J) - A^2  \nonumber \\
 & & \pm
\sqrt{(2(C_2 K + M_2 J) - A^2)^2 + 4 (M_2^2 - C_2^2)(K^2 - J^2 - A^2 k_y^2)} ],  \nonumber \\
K &\equiv& C_0 + C_1 k_z^2 + C_2 k_y^2 - E, \: \: \: \: J \equiv -M_1(k_z^2 - m) - M_2 k_y^2. \nonumber
\end{eqnarray}
The above equation is equivalent to equation~(25) in Ref.~\onlinecite{Gorbar2015} when $k_x$ and $k_y$ are interchanged. The solution of
equation~(\ref{eq:Q}) indicates that the top surface states exist only on the $k_y < 0$ side ($k_{x^{\prime}} > 0$ side). Similarly,
the bottom surface states can be obtained by considering that $p_1$ and $p_2 < 0$ in equation~(\ref{eq:Q}). Then we find that the
bottom surface states appear only on the $k_y > 0$ side ($k_{x^{\prime}} < 0$ side). For the up-sector, the bottom and top surface
states exist on the opposite sides of the $k_z$ axis ($k_{y^{\prime}}$ axis). The calculated CECs of the Fermi-arc surface states
at $E=0$ from this model are shown as black curves in Fig.~\ref{fig:model}c.

Now in the presence of the asymmetric charge transfer at the top surface, a new surface Hamiltonian $\bar{{\cal H}}^d_{s}$ can
be written as ${\cal H}^d_s + \text{diag}(-V,0)$:
\[ \left( \begin{array}{cc}
C_0 + C_1 k_z^2 + C_2 (-\partial^2_x + k_y^2) - A k_y - V      &   -M_1(k_z^2 - m) - M_2 (-\partial^2_x + k_y^2) + A \partial_x   \\
-M_1(k_z^2 - m) - M_2 (-\partial^2_x + k_y^2) - A \partial_x  &  C_0 + C_1 k_z^2 + C_2 (-\partial^2_x + k_y^2) + A k_y
\end{array} \right) . \]
Here the new term $V > 0$ accounts for the charge transfer to the top surface. The small change in the surface Hamiltonian
results in solving equation~(\ref{eq:Q}) with different definitions of $Q_i$ and $p_{1,2}$:
\begin{eqnarray}
Q_i & \equiv & \frac{C_0 + C_1 k_z^2 + C_2 (-p_i^2 + k_y^2) - A k_y - V - E}{M_1(k_z^2 - m) + M_2(-p_i^2 + k_y^2) + A p_i}, \\
p^2_{1,2} &\equiv& \frac{1}{2(C_2^2 - M_2^2)} [ 2(C_2 K + M_2 J) - A^2  - C_2 V \nonumber \\
 & & \pm
\sqrt{(2(C_2 K + M_2 J) - A^2 - C_2 V)^2 + 4 (M_2^2 - C_2^2)(K^2 - J^2 - A^2 k_y^2 - A k_y V - K V )} ].  \nonumber
\end{eqnarray}
The calculated CECs of the surface states at $E=0$ from this model are shown as black and red curves in Fig.~4d.

\vspace{1cm}
{\noindent {\bf Acknowledgments}}\\
J.W.V. and K.P. were supported by the U.S. National Science Foundation grant No DMR-1206354.
The computational support was provided by SDSC under DMR060009N and VT ARC computer clusters.
\\
\\
{\noindent {\bf Author Contributions}}\\
J.W.V. and K.P. performed calculations by using DFT and the effective model.
All of the authors participated in discussion and writing the manuscript.
\\
\\
{\noindent {\bf Competing Financial Interests}}\\
The authors declare no competing financial interests.

\end{document}